\documentclass[prl,aps,twocolumn,floats,superscriptaddress]{revtex4}
\usepackage{graphicx}
\newcommand{\nin}{\noindent}

\newcommand{\be}{\begin{equation}}
\newcommand{\ee}{\end{equation}}
\newcommand{\bea}{\begin{eqnarray}}
\newcommand{\eea}{\end{eqnarray}}

\newcommand{\lb}[1]{\label{#1}}
\renewcommand{\r}{{\bf r}}
\renewcommand{\k}{{\bf k}}
\newcommand{\q}{{\bf q}}

\renewcommand{\v}{{\tilde v}}
\newcommand{\lp}{\left(}
\newcommand{\rp}{\right)}
\newcommand{\la}{\left<}
\newcommand{\ra}{\right>}

\newcommand{\domtD}{\delta \omega \cdot t_D}
\newcommand{\etal}{{\it et al.}}

\newcommand{\ts}[1]{\textstyle{#1}}

\begin{document}
\title{Transverse NMR Relaxation as a Probe of Mesoscopic Structure}
\author{Valerij~G.~Kiselev}
\email{kiselev@ukl.uni-freiburg.de}
\affiliation{Section of Medical Physics, 
University Hospital Freiburg, 
D79106 
Freiburg, Germany}
\author{Dmitry~S.~Novikov}
\affiliation{Department of Physics, 
Massachusetts Institute of Technology, Cambridge, MA 02139}
\date{December 20, 2002}

\begin{abstract}

Transverse NMR relaxation in a macroscopic sample 
is shown to be extremely sensitive to the structure
of mesoscopic magnetic susceptibility variations.
Such a sensitivity is proposed as
a novel kind of contrast in the NMR measurements. 
For suspensions of arbitrary shaped
paramagnetic objects, the transverse relaxation is found
in the case of a small dephasing effect of an individual object.
Strong relaxation rate dependence on the objects' shape 
agrees with experiments on whole blood. 
Demonstrated structure sensitivity is a generic effect that 
arises in NMR relaxation in porous media, biological systems, 
as well as in kinetics of diffusion limited reactions.

\end{abstract}

\maketitle


\nin
NMR as a structure probe is utilized in the fields as diverse 
as chemistry, materials science, geology and biomedicine. 
Pristine specimen found in nature,  
such as rocks or biological tissues, possess complex structure 
at a mesocopic scale. This structure is of primary interest in 
numerous applications. For example, rock porosity in geology  is 
important to assess the oil basin quality.
In biological tissues the mesoscopic scale is set by 
the size of cells and blood vessels whose properties 
carry significant diagnostic and physiological information. 

It is the NMR monitored {\sl diffusion} that is commonly accepted as 
a probe of mesoscopic structure in both inorganic \cite{Mitra92} 
and living \cite{Yablonskiy02,LeBihanbook} specimen. 
In the present Letter we propose a {\sl magnetic susceptibility contrast}
as a structure probe. 
Susceptibility inhomogeneities are often connected to the geometric structure,   
such as pore walls in porous media.
In biological tissues they are brought by paramagnetic cells, such as
deoxygenated red blood cells (RBCs) and iron-enriched
cells in the brain gray matter. 
In some cases the susceptibility 
contrast can be artificially manipulated.

Below we consider NMR 
signal from a suspension of 
arbitrarily shaped weakly paramagnetic objects. 
We demonstrate a significant
individual object shape dependence of the transverse relaxation rate. 
We discuss this result in the biomedical context.
Applications of the biomedical NMR imaging (MRI) are limited
by a spatial resolution $\sim$ 1 mm, which is larger than 
the cell size by 2-3 orders of magnitude. 
Direct resolution enhancement is unfeasible
since today MRI hardware hits physiological limits. 
Our results suggest that further progress can be made by 
a deeper analysis of the NMR signal 
since it contains significant information about the paramagnetic tissue structure
at the scale of several $\mu$m.

We compare our results with experiments on whole 
blood \cite{Brooks95,Spees01,Gomori87},
with the objects being paramagnetic RBCs. 
Previous theoretical efforts in this context 
were focused on the  
effect of paramagnetic inclusions of specific geometries
(spheres \cite{Yablonskiy94,JensenChandra} 
or cylinders simulating blood vessels
\cite{Yablonskiy94,Kiselev98,Bauer99}). 
The effect of object shape was not studied theoretically
although experiments \cite{YeAllen,Gillis95} 
and the Monte Carlo simulation \cite{Gillis95} 
indicate a strong shape dependence of the transverse relaxation.

We model the medium by a suspension of 
$N\!\gg\!1$ identical mesoscopic paramagnetic objects
which are randomly placed and oriented. 
The NMR signal is acquired from nuclear spins 
that freely diffuse in the solvent and in the objects. 
A macroscopic volume $V$ of the suspension 
is characterized by the volume fraction $\zeta = N v_0/V$
of objects ($v_0=$ single object's volume). 
The case of different object species 
is easily accounted for when $\zeta \ll 1$, since they 
contribute additively to the relaxation rate \cite{Yablonskiy94,Kiselev98}.

Transverse relaxation occurs due to two
different mechanisms: 
(i) microscopic spin-spin interactions
at the molecular level, 
(ii) diffusion of spins in the magnetic field induced by 
mesoscopic objects.
Fast processes (i) average out to produce a monoexponential relaxation.
Processes (ii)
are described in terms of 
the transverse magnetization density $\psi(\r)$,
which evolves due to molecular diffusion 
and spin precession with the local Larmor frequency 
varying in space. 
It obeys the Bloch-Torrey equation \cite{Torrey} 
\be                                            \lb{BT}
{\partial \psi \over \partial t}
= \lp D\nabla^2 
- {\frac1{T_2}}
- i\omega_{L}
- i\omega(\r) \rp \psi \ ,
\ee
where $D$ is the diffusion coefficient of the 
molecules that carry spins, and $T_2$ is the relaxation time due to 
the microscopic interactions. The relaxation rate $1/T_2$  
is insensitive to magnetic field inhomogeneities at the
mesoscopic scale. Rather, it characterizes local chemical
composition. We assume that $D$ and $T_2$ 
are the same inside the objects and in the solvent.
The constant term $\omega_{L}$ provides the Larmor precession 
in the homogeneous main field, and  
$\omega(\r) = \sum_{n=1}^N \omega_0(\r-\r_n)$  is the deviation 
from $\omega_L$ due to the local magnetic
fields induced 
by randomly located paramagnetic objects (as described later).

The signal $S(t)$ from a macroscopic sample 
is the sum of all spin magnetic moments 
regardless of their initial positions and 
their Brownian trajectories after the excitation \cite{Kiselev98}.
In terms of the Green's function
$\psi(\r,\r_0,t)$ of (\ref{BT}),
defined by the initial condition 
$\psi(\r_0,\r, \ t\!=\!0) = \delta(\r-\r_0)$,
\be                                               \label{def-S}
S(t)  =  {\frac1V}\!\!\int\!\! d^3\r  \, d^3\r_0 \; \psi(\r,\r_0,t) 
\,\equiv e^{-i\omega_{\! L} t -t/T_2} \; s(t)\ . 
\ee
Microscopic processes decouple due to Eq.~(\ref{BT}): 
$\psi \!=\! e^{-i\omega_{\! L} t - t/T_2} \, \phi$,
with  $\phi(\r,\r_0,t)$ accumulating {\it mesoscopic} effects. 
The corresponding signal attenuation factor
\be                                               \label{def-s}
s(t) = {\frac1V} \int\!\! d^3\r  \, d^3\r_0 \; \phi(\r,\r_0,t) \ ,
\quad s(0) = 1 \ ,
\ee
describing these effects is the main object of our focus. 

Consider the mesoscopic part 
$M(\r_0, t) \!=\! \int\! d^3\r \,\phi(\r_0,\r,t)$
of the spin packet magnetization. 
The $d^3\r_0$ integration 
in (\ref{def-s}) 
effectively averages $M(\r_0, t)$ over randomly positioned objects. 
For $\zeta\!\ll \!1$, $M$  is a product 
of factors contributed by individual objects \cite{Kiselev98}. 
In this case $s(t)$ is expressible in terms 
of a {\it single} object dephasing effect $f(t)$ 
\cite{Yablonskiy94,Kiselev98}:
\be                                              \label{def-f}
s  = e^{-\zeta f} , \ \ \
f(t) = \la \!\int \! {d^3\r_0 \over v_0}
\lp
1 - \int \! d^3\r \,\eta(\r_0,\r, t)
\rp \ra_{o} .
\ee
Here $\eta$ is the mesoscopic part of the spin packet 
magnetization density in the presence of a single object, 
\be                                      \label{BT-reduced}
{\partial \eta \over \partial t}\!
= \!\left( D\nabla^2 - i\omega_0(\r) \right) \eta \ , \quad
\eta(\r_0,\r, t\!=\!0)\!=\!\delta(\r  - \r_0) \ ,
\ee
and average $\la\ \ra_{\! o}$ in (\ref{def-f}) is taken over the
object's orientations.

In the main field $B_0 {\bf\hat{z}}$ 
each paramagnetic object induces a local Larmor frequency shift  
$\omega_0(\r)$ that is determined by the object's 
{\sl susceptibility profile} $\chi(\r)$. 
Below we use uniformly magnetized objects to compare with experiments:  
$\chi(\r)\! =\! \chi\!\cdot\! v(\r)$, $\chi\!\ll\!1$, where
$v(\r)$ is a shape function: $v\!=\!1$ inside and $v\!=\!0$ outside
the object. 
A convolution in $\r$,~ $\omega_0$ in the Fourier space is 
\be						\label{def-omega}
\omega_0(\k) = \delta \omega \cdot Y(\hat\k) \cdot  \v(\k) \ , 
\quad \delta \omega = 4\pi\chi\omega_L \ , \\ 
\ee
where $Y(\hat \k) = 1/3-k_z^2/k^2$ 
is the longitudinal projection
of an elementary magnetic dipole field, and 
the object's form factor $\v(\k)$ is the Fourier transform of
$v(\r)$.

Transverse relaxation is qualitatively different  
in the limits of strong and weak dephasing.
Introduce effective object radius $\rho$ 
as that of a sphere of a volume $v_0$.
Water molecules pass by the object during the diffusion time 
\be \label{def-tD}
t_D = {\rho^2 \over D} \ , \ \ \ {\rm where} \ \ 
\ts{4 \over 3}\pi\rho^3 \equiv v_0 = \int \! d^3\r \, v(\r) \ .
\ee
Typical phase acquired by the spins is $\domtD$.
In the pre\-sent work we focus on a weak dephasing case 
$\delta\omega\!\cdot\!t_D \!\ll\! 1$ 
(diffusion narrowing regime). 
This regime covers a variety of experiments, 
in particular spin dephasing in diamagnetic and paramagnetic  
samples in the field $B_0 \!\lesssim\!1$T. 

We find the Green's function $\eta$ of Eq.~(\ref{BT-reduced}) 
perturbatively in the small parameter $\domtD$,
and use Eq.~(\ref{def-f}) to obtain $f(t)$.
This approach is analogous to the Born series 
for the quantum mechanical scattering amplitude. 

\begin{figure}[b]
\includegraphics{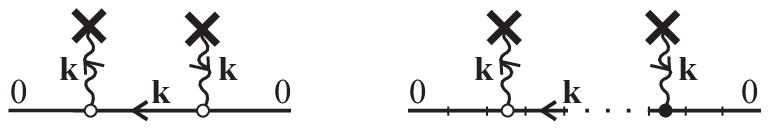}
\caption[]{\label{figdiagrams}
Second order processes for $f(t)$.
{\it Left:} FID relaxation.
Circles, wavy lines and crosses stand for 
$-i\delta\omega$, $Y(\k)$, and $\v(\k)$ respectively. 
Solid lines represent free propagators $\eta^{(0)}(\k, \tau)$
in time intervals between interactions. 
External momenta are set to zero due to Eq.~(\ref{def-s}). 
{\it Right:} CPMG relaxation.
Each section represents a free propagator $\eta^{(0)}$
in the interval $\Delta t$ between successive refocusing pulses.
Complex conjugation on every other interval $\Delta t$ 
is indicated with the filled circle.
Equation (\ref{g-cpmg}) is obtained as a sum of all such configurations. 
}
\end{figure}

The zeroth order in $\domtD$ describes free diffusion. 
In this case the total magnetization of each spin packet is conserved, 
$\int \! d^3\r \; \eta(\r_0,\r; \, t) = 1$ in (\ref{def-f}),  
and $s(t) = 1$.
The first order correction to $f$ vanishes 
since it is proportional
to the angular average of the dipole field. 
The expression for $f$ is dominated by the second order in $\domtD$
(Fig.~\ref{figdiagrams}):
\bea                                              \lb{basicresult}
f(\tau) =
{2\pi \alpha^2\over 15}
\int_0^{\infty} \!\! {dq \over q^2} \, g (q^2 \tau)
\int \!\! {d{\bf \hat q}\over 4\pi} \, 
\left|
{\v(\q)\over \v({\bf 0})}
\right|^2 \ , 
\\ 						  \lb{defalpha}
\alpha \equiv \ts{2\over 3\pi} \ \domtD \ .  
\eea 
Here $\tau$ is the dimensionless time $\tau=t/t_D$.  
The inner integral in Eq.~(\ref{basicresult}), 
which is taken over the directions of $\q = \k \rho$,
depends exclusively on the object shape. 
The object size enters Eq.~(\ref{basicresult})
only through the diffusion time $t_D$, Eq.~(\ref{def-tD}).  
The function $g$ depends on the particular sequence of the 
radiofrequency (rf) pulses applied to manipulate the spins
and is discussed later.

As a conservative estimate, 
the formal series for $f(t)$ converges when $\alpha < 1$.
Indeed, each successive term in the perturbative expansion of $f(t)$
is multiplied by the dimensionless factor
$i\alpha = i\, 4\pi/(2\pi)^3 \, (\domtD) \, (v_0/\rho^3)$.
Angular integrations of the products $Y(\hat{\k}_1)Y(\hat{\k}_2)...$
improve convergence by bringing additional factors $<1$ that are
object shape specific and hard to account for in general.

Odd orders of the expansion in $\alpha$ are imaginary. 
They renormalize the homogeneous component
of the suspension's magnetic susceptibility. 
Since the first order vanishes the correction to $\omega_L$
is proportional to $\alpha^3$.
The signal attenuation is determined by the even orders in $\alpha$.
The correction to (\ref{basicresult}) is of the order
of $\alpha^4$ and is negative.

Consider the free induction decay (FID), an evolution
after a single rf $\pi/2$ pulse 
which creates the maximal transverse spin magnetization. 
The function $g$ in (\ref{basicresult}), 
denoted as $g_{\rm FID}$, 
is proportional to a time convolution of the three
free diffusion propagators 
$\eta^{(0)}(\q,\tau)=\theta(\tau)e^{-q^2\tau}$, 
$\ \theta(\tau)$ being a unit step function
(Fig.~\ref{figdiagrams}, left):
\be                                             \lb{g-fid}
g_{\rm FID}(q^2 \tau)
=
q^2 \tau - 1 + e^{-q^2 \tau} .
\ee

To reduce sensitivity to large scale field inhomogeneities, 
samples are often irradiated by a number of refocusing rf $\pi$ pulses. 
Each such pulse quickly 
rotates the spins by $\pi$ around an axis which is 
transverse to $\hat{\bf z}$. 
This is equivalent to a complex conjugation of $\eta$ developed 
up until this time moment. The resulting distribution $\eta^*$ is the 
initial condition for the further evolution. 

In the spin echo (SE) technique \cite{SE-Hahn} a single $\pi$ pulse 
is applied at the time $t_E/2$ and the signal is measured at $t=t_E$. 
The corresponding $g$ function reads
\be                                                \lb{g-se}
g_{\rm SE}
=q^2 \tau_E - 3 + 4e^{-q^2 \tau_E/2} - e^{-q^2 \tau_E} \ ,
\quad \tau_E={t_E\over t_D} \ .
\ee
In the CPMG (Carr-Purcell-Meiboom-Gill) protocol \cite{CPMG}
refocusing $\pi$ pulses are generated in a long train and the steady
state signal is studied as a function of the interpulse interval $\Delta t$
(Fig.~\ref{figdiagrams}, right):
\be                                                  \lb{g-cpmg}
g_{\rm CPMG}
=q^2 \tau - 2\tanh{\frac{q^2 \tau}2} \ ,
\quad \tau={\Delta t\over t_D} \ .
\ee

\begin{figure}
\hbox to \columnwidth
{
\includegraphics[height=1.2in]{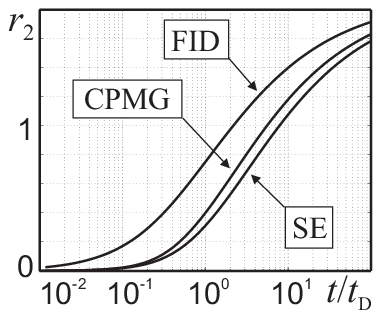}
\hfil  
\includegraphics[height=1.2in]{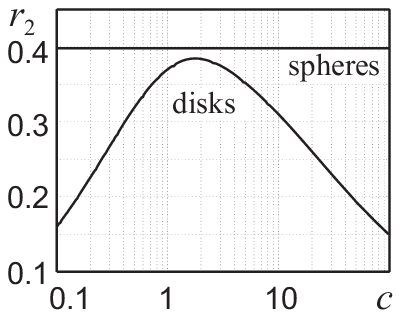}
}
\caption[]{\label{figsignal}
Mesoscopic relaxivity 
$r_2\!=f(\tau)/\tau$ 
for $\alpha^2\!=\!15/2\pi$. 
{\it Left:} 
Objects are spheres, 
$t\!=\!t_E$ for the SE, $t\!=\!\Delta t$ for the CPMG. 
{\it Right:} Shape effect: disks {\sl vs.} spheres. 
CPMG relaxivity $r_2(\Delta t/t_D \!=\! 1)$,
objects are disks with height-to-radius
ratio $c$, and spheres of the same volume.
}    
\end{figure}

Equations (\ref{g-fid})--(\ref{g-cpmg}) yield that at $\tau \ll 1$, 
$f \propto \tau^2$ for the FID and 
$f \propto \tau^3$ for the SE and the CPMG sequences.  
Asymptotic expansion of (\ref{basicresult})
in $\tau^{-1/2}$ at $\tau\gg 1$ gives 
\be                                              \lb{asylong}
r_2 \equiv {f(\tau)\over \tau} \simeq
{2\pi \alpha^2\over 15}\!
\left(
\int_0^{\infty} \!\!\!\! dq
\int \! {d{\hat \q}\over 4\pi} \left| {\v(\q)\over \v({\bf 0})}  \right|^2
- {A\over \sqrt{\tau}}
\right) \ ,
\ee
with 
$A_{\rm FID}=\sqrt{\pi}$, $A_{\rm SE}=(2\sqrt{2}-1)\sqrt{\pi}$, 
and $A_{\rm CPMG}=(2\sqrt{2}-1)\,\zeta(3/2)\,/\sqrt{\pi} \approx 2.695$
for the considered pulse sequences.
The dimensionless NMR relaxivity $r_2$
is shown in Fig.~\ref{figsignal}, left, for the case of the homogeneously
magnetized spherical particles. 
Shape dependence is illustrated in Fig.~\ref{figsignal}, right,
for the case of disk-shaped objects.
The height-to-radius ratio $c$ 
defines the disk shape, with $c=0.5$
being close to the intact RBC.


Below we analyze our results, Eqs.~(\ref{basicresult}) and (\ref{asylong}).

(i) 
{\sl Relaxation (\ref{def-f}) crucially depends on the shape of the object.}
It is the form factor $\v(\k)$ that 
governs the convergence of the integral for large $\q=\k \rho$ 
in (\ref{basicresult}). The integral converges at $k\sim 1/\rho$, 
allowing one to probe the object's structure. 
(A quantum mechanical analogy
is scattering amplitude dependence on the 
form factor of the external potential.) 
A point-like magnetization $v\!\propto\!\delta(\r)$ 
causes a divergence in Eqs.~(\ref{basicresult}) and (\ref{asylong}).
In the present case this ``nonrenormalizability'' 
(non-universal cutoff dependence) 
effectively increases sensitivity 
in the NMR measurements.

(ii)
{\sl Shape sensitivity is a consequence of a {\it singular}
interaction} $Y\!\sim \! r^{-\nu}$ between nuclear spins and objects.
Consider the case when the singularity in $Y$ is cut off at a scale $r<a$.
Then $Y(\k) \to  0$  as $ka > 1$. 
If $a\!>\!\rho$, the integral in Eq.~(\ref{basicresult}) is insensitive to 
the form factor since it converges at $k\!<\!1/a\!<\!1/\rho$, 
destroying shape sensitivity. 
Physically, such a cutoff
introduces a spherical ``cloud'' of a radius $a$ around each object.
This cloud smears information about the object's structure.
The power necessary for shape dependence is $\nu>2+(d-2)/N$ 
for the $N$-th order in $d$ dimensions. 
Thus both magnetic dipole ($\nu = 3$) 
and contact interaction $Y=\delta(\r)$ in $d\!=\!3$ 
yield shape sensitivity already in the second
order, as shown above. 

(iii) 
{\sl Shape sensitivity is present for any field $B_0$.}--
Above we demonstrated shape sensitivity 
in the domain where the perturbative approach is reliable 
($\alpha \!\ll\!1$).
We now prove it for any $\alpha$.
Integrals such as (\ref{basicresult}) whose convergence is form factor dependent 
appear in each order of the perturbation series for $f(t)$. 
Although angular integrations impede explicit 
summation of this series, they 
do not cause nonanalyticity at $\alpha\!=\!0$, and thus
radius of convergence in $\alpha\!\propto\! B_0$ is finite. 
Therefore the series 
can be analytically continued to the large field domain $\alpha > 1$ 
where the perturbation theory formally breaks up. 
The final result for $f(t)$ would still be form factor dependent,
which proves shape sensitivity for $any$ field.

(iv) 
{\sl Shape sensitivity is a generic effect}.--
Consider {\sl a diffusion limited chemical reaction}
on impurities with a shape $u(\r)$.
The FID signal analytically continued by $-i\omega(\r) \to u(\r)$
gives the impurity shape dependent reaction rate.

\begin{figure}[t]
\hbox to \columnwidth
{
\includegraphics[height=1.2in]{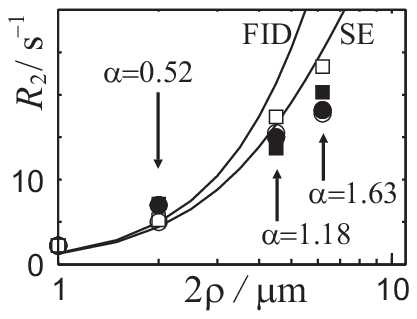}
\hfil 
\includegraphics[height=1.2in]{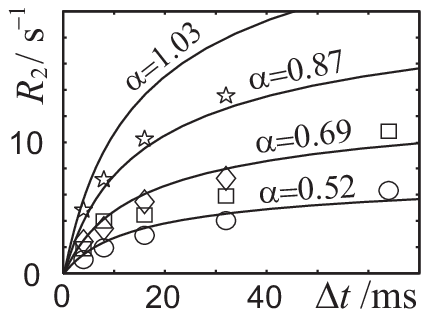}
}
\caption[]{\label{figexperiments}
Theory (lines) {\sl vs.} experiments (symbols).
{\it Left:} 
Relaxation rate $R_2=-(\ln s)/t$
for the FID (boxes) and SE (circles)
as a function of the particle diameter.
Filled and hollow symbols correspond
to measured and Monte Carlo simulated relaxation rates \cite{Weisskoff94}.
{\it Right:} 
CPMG relaxation rate for the human blood samples \cite{Gomori87}
for $B_0=1.41, 1.18, 0.94, 0.71$~T (from top to bottom). 
Experimental errors are $10$--$20\%$ \cite{Gomori87}.
Following our discussion after Eq.\ (\ref{basicresult}),
theory agrees with experiment 
for small $\alpha$ and overestimates it for $\alpha\simeq 1$. 
}    
\end{figure}

Below we compare the results (\ref{basicresult}) and (\ref{asylong})
with experiments. 
As a test we use the reported relaxation rate 
in dilute ($\zeta=0.02$) suspensions of polystyrene microspheres in 
paramagnetically doped water \cite{Weisskoff94}
(Fig.~\ref{figexperiments}, left).

Further experiments were performed on
the deoxygenated blood with a high RBC volume
fraction $\zeta=0.40-0.60$ \cite{Brooks95,Spees01,Gomori87}.
To apply Eqs.~(\ref{basicresult}) and (\ref{asylong}) one needs to 
take into account a slower diffusion inside the cells and 
to extend our approach for large $\zeta$.
The former will be considered elsewhere. 
For now, we obtain
upper and lower estimates for the relaxation rate by using 
in Eq.~(\ref{basicresult}) the values 
$D_{\rm in}$, $D_{\rm out}$
of the diffusion coefficient 
in erythrocytes and plasma respectively. 

The $\zeta \!\sim \!1$ case poses a challenging task
equivalent to finding the statistical
sum of a dense gas of objects. 
Instead we replace $\zeta$ by $\zeta(1-\zeta)$ in (\ref{def-f}),
which is well supported experimentally \cite{Thulborn82}.
Such a replacement is justified by the virial expansion. 
Eq.~(\ref{def-f}) treats exactly the first cumulant 
of the statistical sum. The second cumulant provides 
an ${\cal O}(\zeta^2)$ {\it negative} correction.
This together with a vanishing mesoscopic contribution as $\zeta\!\to \!1$  
justifies a quadratic polynomial interpolation, $\zeta(1-\zeta)$.
The latter is correct as
$\zeta\!\to\! 0$ and $1$, and describes a crossover between the 
dilute and the extremely dense cases.


The relaxation rate in deoxygenated blood measured in \cite{Brooks95}
quadratically depends on the magnetic field, 
$-(\ln s)/t = \kappa_1 B_0^2$, 
in agreement with Eq.~(\ref{basicresult}). 
The proportionality coefficient $\kappa_1$
was found to be $7.2~{\rm s^{-1}T^{-2}}$
for the CPMG pulse sequence with $\Delta t=4$~ms.
In \cite{Brooks95}, the field range $B_0 = 0.05 - 1.5$~T 
yields $\alpha = 0.033 - 0.99$. 
We calculated $\zeta = 0.55$ from the parameters given in \cite{Brooks95},
utilized the magnetic susceptibility of the deoxygenated
RBCs $\chi=2.7\times 10^{-7}$ \cite{Spees01},
and simulated the intact erythrocytes by disks of the known volume
of $87~{\rm \mu m^3}$ with the height-to-radius ratio of $c=0.5$. 
Using $D_{\rm out}=2.20~{\rm \mu m^2/ms}$ and $D_{\rm in}=0.76~{\rm \mu m^2/ms}$ 
\cite{Li98}, our theory gives $4.7 < \kappa_{1\rm th} < 5.6~{\rm s^{-1}T^{-2}}$.

To assess this result we note that neither the 
susceptibility of RBCs nor their actual shape
was reported in \cite{Brooks95}. Chemicals
used to treat the samples are likely to change osmotic pressure
in plasma, which would deform the RBCs
thus changing all relevant parameters. 

Quadratic dependence of the SE blood relaxation rate on $\chi$, 
which follows from Eq.~(\ref{basicresult}), was confirmed 
by varying the RBC oxygen saturation $y$ in the field $B_0=1.5$~T
\cite{Spees01}: 
$-(\ln s)/t = \kappa_2 (0.95-y)^2$, with the measured coefficient
$\kappa_2= 55 \ {\rm s^{-1}}$ for $\zeta=0.3$ and 
$\kappa_2= 59 \ {\rm s^{-1}}$ for $\zeta=0.4$. 
Our approach results in the corresponding ranges 
$26 < \kappa_{2\rm th} < 56 {\rm \ s^{-1}}$ and 
$30 < \kappa_{2\rm th} < 64 {\rm \ s^{-1}}$.  

The CPMG relaxation rate $R_2=-(\ln s)/t$ 
in the whole blood was measured \cite{Gomori87} 
as a function of the interecho interval 
(Fig~\ref{figexperiments}, right) 
for $0.71\!<\!B_0\!<\!1.41$~T. 
We simulated blood as described above 
using $D\!=\!D_{\rm out}$ for plasma.
The use of the value $D_{\rm in}$ instead of $D_{\rm out}$
yields about the same rate $R_2$
for the short times and approximately a twofold increase of $R_2$ for
the large times. 

This brief survey shows that, although crude, our model captures
essential features of the NMR relaxation.
Experiments at higher fields \cite{YeAllen,Gillis95}
confirm the shape dependence for $\alpha>1$. 
Their results can be well described by adjusting 
$t_D$ and $\alpha$ \cite{JensenChandra} or by fitting to a simple 
chemical exchange model \cite{Brooks01}.
However, fitting has a predictive power when the signal universally 
depends on a handful of phenomenological parameters.
Shape sensitivity makes such a fitting meaningless in the case of 
varying tissue structure. Because of the same reason, in experiments
analogous to \cite{Brooks95,Spees01,Gomori87,YeAllen,Gillis95} 
it is essential to control 
volume fraction, shape and susceptibility of paramagnetic
objects, and effective diffusion coefficient in the sample.


To conclude, we showed that transverse relaxation from a suspension of paramagnetic
objects is extremely sensitive to the shape of 
the individual object. This sensitivity 
to geometric structure is a 
generic effect that can be employed 
as a novel type of contrast in NMR measurements thus effectively increasing
spatial resolution.

D.N. appreciates a visiting grant and
hospitality of the Section of Medical Physics,
University Hospital Freiburg.


\end{document}